\RequirePackage[hyphens]{url}

\documentclass[final,3p,times,twocolumn]{elsarticle}

\usepackage{amssymb}

\usepackage{url}
\usepackage{xcolor}
\usepackage{subfigure}
\usepackage{booktabs}

\PassOptionsToPackage{hyphens}{url}\usepackage{hyperref}

\journal{Computer Networks}

\begin{document}

\begin{frontmatter}



\title{A measurement-based approach to analyze the power consumption of the softwarized 5G core}


\author[UniTn,Athonet]{Arturo Bellin}
\author[UniTn]{Fabrizio Granelli}
\author[Athonet]{Daniele Munaretto}

\affiliation[UniTn]{organization={Department of Information Engineering and Computer Science (DISI), University of Trento, \& C.N.I.T.},
            city={Trento},
            postcode={38123}, 
            state={},
            country={Italy}}

\affiliation[Athonet]{organization={Research and Innovation Department, Athonet a HPE acquisition},
            addressline={via Cà del Luogo 6/8}, 
            city={Bolzano Vicentino},
            postcode={36050}, 
            state={},
            country={Italy}}

\begin{abstract}

In light of the ever growing energy needs of the ICT sector, a value that is becoming increasingly important for a mobile network is its \emph{power consumption}. However, the transition away from legacy network deployments tightly coupled with the underlying hardware and the adoption of the Network Function Virtualization (NFV) paradigm has made more difficult to accurately evaluate their energy and carbon footprint.
In this paper, we propose and validate a measurement-based approach to analyze the power consumption of a virtualized 5G core network (5GC) deployment. We design an experimental testbed using commercial off-the-shelf (COTS) hardware and open-source software as a sample architecture simulating an edge computing node and supporting three different virtualization options. 
We make use of both hardware-based and software-based power meters to investigate the power consumption trends associated with increasing levels of traffic and multiple 5GC deployment types.
The results show the feasibility of a real-time power monitoring system and highlight how deployment choices, such as virtualization framework and 5GC software, can significantly impact on the power consumption of the network.

\end{abstract}



\begin{keyword}
5G core network \sep power consumption \sep NFV \sep edge computing



\end{keyword}

\end{frontmatter}


\section{Introduction}
\label{sec:Introduction} 

The widespread deployment of 5G is expected to offer a modern and flexible network infrastructure to support novel services and applications. Indeed, this new mobile network architecture will integrate several novel concepts, including softwarization and virtualization of the network core, advanced radio interfaces, edge computing and private mobile networks. 
The exponential growth of Internet traffic (increasingly mobile) and increase in users and "Things" to be connected will require the adoption of scalable concepts within the design of 5G networks, focusing on sustainability and careful control of the consumed power. 

Traditionally, the attention of manufacturers was mainly focused on the Radio Access Network (RAN) \cite{ericsson:ran} due to the complexity of the base stations and their increased number due to the continuous reduction of the mobile cell size. 
Nevertheless, the impact of the innovative design of the 5GC is not yet fully analyzed, also due to its virtualized nature.
In fact, the 5GC Service-Based Architecture (SBA) fully embraced the paradigm of NFV, allowing for the deployment of network functions on virtualized COTS hardware or on public cloud computing infrastructures instead of application-specific integrated circuit.

This allows for a much easier dynamic and distributed deployment of virtual network functions (VNFs) within the network infrastructure, enabling new use cases and providing better privacy, latency and reliability (e.g. through network slicing).
Moreover, private 5GCs entirely or partially deployed on-premise are more likely to leverage smaller and less efficient COTS hardware, without the scaling and consolidation benefits provided by larger data centers.
According to the Natural Resources Defense Council report \cite{NRDC:datacenter}, large hyperscalers server infrastructure represented less than 5\% of the United States’ data center energy use, with the remaining 95\% consumed by less efficient small and medium data centers.
The ITU-T has recently defined a key performance indicator (KPI) for the carbon emission intensity of a network focused on the energy consumption with respect to served data traffic, not only encouraging the reduction of network electricity consumption, but also advocating the use of low-carbon energy supply and the improvement of energy utilization efficiency \cite{ITU:carbon}.
Based on the above considerations, being able to consistently measure and characterize the power consumption of the 5GC represents a relevant topic, as a tool to apply proper operations, administration and management (OAM) decisions to enable 5G to be sustainable and scalable according to the Zero-touch network and Service Management (ZSM) paradigm \cite{ETSI:ZSM}.

In this paper, we focus on the characterization of the power consumption of the core of 5G systems (5GSs), that is, the 5GC. 
Modeling the power consumption of the 5GC, given the heterogeneity of the underlying hardware and virtualization technologies, is not simple. Only a few works are available in the literature on this subject.
As we demonstrate in this paper, the choice of both hardware and hypervisor strongly influences the overall energy demand.
It is not possible to rely only on fixed models and technical specifications from the hardware manufacturers since different networks, or different components within the same network, may be deployed on diverse kinds of infrastructures. 

\subsection{Paper contributions and structure}

In this context, the main contributions of our work are as follows: 
\begin{itemize}
    \item We present a proof-of-concept for a real-time monitoring system for the power consumption of a 5GC, outlining how it can be used to orchestrate VNFs at the network edge. See subsection \ref{subsec:methodology}.
    \item We investigate the power consumption trends of open-source implementations of a 5GC instantiated on COTS hardware, analyzing different deployment options. See section \ref{sec:Experimental_results}.
    \item We compare and model the measurements obtained using hardware-based and software-based power meters discussing the advantages and challenges of both. See subsection \ref{subsec:Virtualization_comparison}
    \item We make the collected power consumption data from all evaluated environments available freely on web, so that it can be used by the scientific community in subsequent studies or as a basis for comparison. See section \ref{sec:data}.
\end{itemize}

The structure of the paper is as follows: Section \ref{sec:Related_work} provides a review of existing works related to power consumption in virtualized systems and 5GC, while Section \ref{sec:proposed_methodology} describes the proposed methodology and measured KPIs. Experimental results are presented and analyzed in Section \ref{sec:Experimental_results}. Finally, conclusions and outlines about future work on the topic are provided by Section \ref{sec:Conclusions}, which concludes the paper.

\section{Related work}
\label{sec:Related_work}

Several studies available in the literature have analyzed the power consumption of cloud infrastructures and data centers, but it is not clear how this knowledge can be translated to the telco world and especially to edge 5GC deployments running on COTS hardware.
One of the most exhaustive and up-to-date survey on the topic is \cite{Depasquale2023}, where authors summarize the dynamics and the results of recent research efforts in power consumption measurement and power models of virtual entities in the telco cloud sector.
Study in \cite{Jiang2019} presents a comprehensive examination of the power and energy usage of four of the most used hypervisors and one container engine by testing them on six different hardware platforms, including a variety of rack server structures, one desktop server, and one laptop. Power measurements are gathered during different levels of computation-intensive, memory-intensive, and mixed Web server-database workloads. The findings highlight the different characteristics of each hypervisor and their better suitability for specific workloads or platform with none significantly surpassing the others in performance.
A similar comparison is presented in \cite{Morabito2015} with the aim of characterizing the power consumption of different virtualization technologies in the idle state and under CPU, memory, and networking intensive workloads.
Based on the experimental measurements, the authors conclude that, for the analyzed solutions, significant differences are present only during the networking stress test where container solutions perform better than hypervisors.
Further confirmation of these results is presented in \cite{Shea2014}, where hardware virtualization and paravirtualization solutions proved to be up to 40\% more energy demanding than a standard bare-metal machine performing the same networking tasks. On the other hand, the performance and power consumption of container
virtualization systems is comparable to the bare-metal.
While these works present very useful insight, they do not consider the specific needs and scenarios typical of a 5GC deployment and of an edge environment.
A similar analysis but more targeted towards 5GC deployments is presented in \cite{Behravesh2019}, in which the authors benchmark instances of an Apache HTTP server and of a Redis server running on VM, containers and unikernels. The work done in \cite{Behravesh2019} was expanded in \cite{Aggarwal2020} with the addition of Kata containers, a virtualization solution that combines benefits of both containers and VMs.
The common conclusion is that the overhead of the virtualization can be quite significant in terms of resource usage, nevertheless each virtualization environment has some advantages making it the preferred solution in some specific use-cases.
Additional considerations on this subject are presented in \cite{Adoga2022} which shows the performance gains from deploying VNFs on heterogeneous frameworks.
Only recently have studies presented performance analysis of actual 5GC instances. A comparison of CPU usage, latency, connection time and other metrics between open-source 5GC solutions are presented in \cite{Reddy2023} and \cite{Lando2023}.

Our previous study \cite{Bellin2023} covers some of the early results and a preliminary assessment of the realistic power consumption of a 5GC deployed in a network edge environment. The main limitation of the study is that the proposed power monitoring system only considers hardware-based power meters and was validated on a single 5GC implementation and a single connected UE.

\subsection{Lessons learned from the literature and motivation of the work}
\label{subsec:Motivation}

Based on the aforementioned literature, we infer that virtualization typically adversely affects performance and energy consumption. However, certain frameworks may offer advantages in terms of agility, isolation, and ease of management for particular use cases and workloads.
The increasing amount of research that has been carried out in recent years prove the relevance of the topic, not only in light of the ever growing environmental concerns regarding the ICT sector but also the desire of mobile network operators to reduce their operational expenditure.
Nonetheless, power consumption monitoring and optimization in mobile networks is still very much an open topic for research with some significant challenges that still have to be tackled.

In addition to previous efforts, our study includes the following novelties:
\textcolor{black}{
\begin{itemize}
    \item Design and implementation of an experimental testbed for the simulation of an edge computing environment using COTS hardware and open-source software.
    \item Acquisition of power measurements using real 5GC instances instead of benchmarking softwares.
    \item Analysis of the usage of software-based and hardware-based power meters.
    \item Providing insights on the consumption of the single VNFs composing the 5GC based on actual testbed measurements.
\end{itemize}
}

\section{The Proposed Methodology}
\label{sec:proposed_methodology}

\subsection{Methodology Description}
\label{subsec:methodology}

\begin{figure}[h]
\centering
\includegraphics[width=\columnwidth]{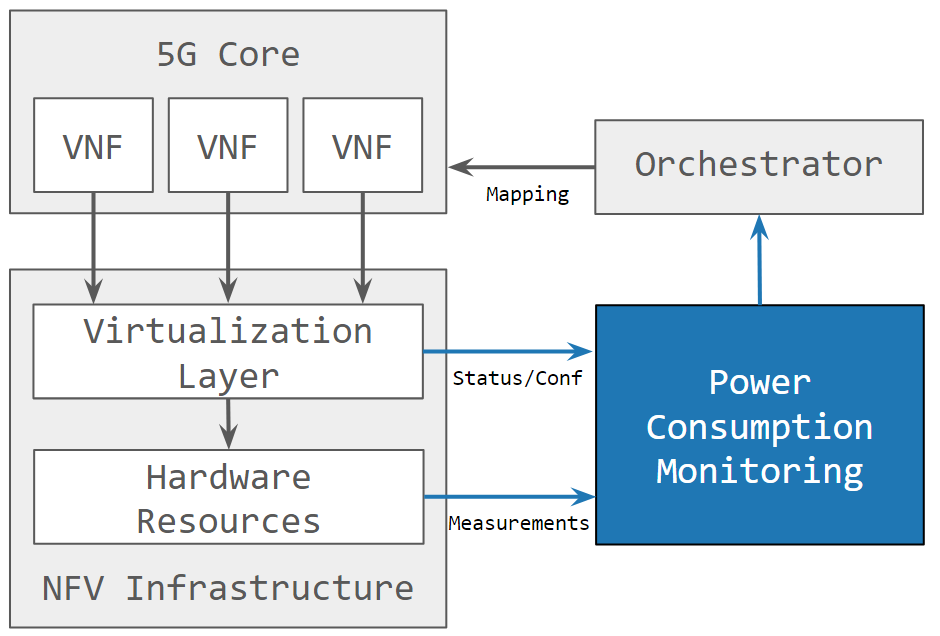}
\caption{Diagram showing a power consumption monitoring system integrated with a virtualized 5GC deployment. The parts we are concerned about in this work are highlighted in blue.}
\label{fig:orchestration_diagram}
\end{figure}

This paper proposes to use a measurement-based approach to analyze the power consumption of the 5GC VNFs. Indeed, the usage of reliable measurements allows to build a clear picture about the association of power to the different functionalities supported by heterogeneous hardware devices and to reduce the complexity of requiring to model each and every single component of the system (which might be often not possible or too complex).
Characterization of power consumption of the 5GC requires the usage of power meters to collect information from the hardware infrastructure, and to combine those with other information of the status of the 5GC (e.g. traffic, CPU usage) in order to define a proper power consumption model to associate to each VNF.
It is important to underline at this point that power consumption can be only directly measured on the hardware. The measured power consumption will thus depend on the specifics of the hardware platform, the chosen virtualization architecture, and the supported VNFs. 
In this regard, the edge computing environment represents a real challenge as it is highly heterogeneous and distributed by nature. Collecting the necessary KPIs in a reliable and consistent manner requires additional efforts as very little standardization is available \cite{ts:28.310}.
It is therefore necessary to make use of dynamic and adaptable mechanisms for the real-time monitoring of the network power consumption, which can be fed to the network orchestrator to enable it to take the appropriate OAM decisions, enhancing it to support energy saving as one of the optimization parameters.
The conceptual integration between this mechanism and a virtualized 5GC deployment is shown in Fig.~\ref{fig:orchestration_diagram}. 
Power measurements are continuously gathered from the hardware infrastructure together with the status and configuration of the virtualization layer. Such information can be processed and forwarded to the network orchestrator which acts on the 5GC configuration and deployment accordingly.
In addition, the metrics can be directly exposed to all involved stakeholders and to the final users to increase the awareness towards the energy consumption and carbon footprint associated with the consumed services.
To this end, there are significant efforts by public and private entities \cite{6Green} towards the definitions of new services and extension of the 5G SBA existing standardized NFs, such as the Network Data Analytics Function and Network Exposure Function, to support energy and carbon metrics observability mechanisms across the entire edge-cloud continuum. 
The introduction of an a SBA exposure layer can provide simple and well-defined intent-based interfaces to enable the coordination of the stakeholders in the allocation of resources and drive sustainable network operations.

\subsection{Power metering solutions}
\label{subsec:Software-based_power_meters}

Historically, the main solution for energy monitoring involve the use of electronic and electromechanical power meters placed between the computing infrastructure and the power source. On the one hand, these meters produce a reliable measure of the overall system consumption and can be used with a great variety of infrastructures. On the other hand, they can often be very expensive and present difficult or proprietary means of accessing the data.
Nowadays, the power consumption of computing units can be measured without the need of an external physical appliance using specific software-based power meters.
These tools enable the collection of power consumption metrics at various levels of detail and granularity, ranging from monitoring a single process to the entire computing unit. This is one of the main advantages compared to physical meters that only provides the overall consumption of the system.
Another advantage is the much easier and faster deployment process, often requiring only the installation of a small script on the host OS. Updating the meters is also more convenient, especially at scale, as it requires little to no financial investment that would otherwise be required to replace the old hardware.
On the other hand, the main drawback of software-based power meters is their limited scope, as they are capable of monitoring only specific components, usually CPU, GPU and RAM. To overcome this limitation analytical models are often used in addition to the low-level measurements compromising the accuracy of the final result. 
Another disadvantage is the more invasive nature compared to external power meters. It is possible, especially when measuring with high polling rate or on low performance hardware, that the measuring software itself requires a significant share of CPU and therefore alter the power consumption data. It is not difficult to account for this additional consumption but one must be careful.

A more thorough analysis of benefit and disadvantages of various software-based power meters can be found in \cite{Jay2023}. 
In addition, the survey in \cite{Ismail2020} presents a comparison of different power models formulas and techniques in a unified testing environment taking into consideration heterogeneous workloads.

\subsection{Sample Architecture}
\label{subsec:Architecture}

This section describes the sample architecture used for laboratory experiments, to demonstrate the proposed methodology.

The testbed comprises 5 Intel\textsuperscript{\tiny\textregistered} NUC units with the following hardware specifications. Three NUCs are equipped with a i5-7260U CPU @2.20GHz, 8GB of RAM, 240GB SSD, 1 Gigabit Ethernet and Dual Band Wireless connectivity. The other two NUCs are equipped with slightly more powerful hardware, in particular a i5-1145G7 CPU @2.60GHz.
All units are running the same version of Ubuntu 20.04 Desktop operating system (OS) and are connected to a five-ports gigabit switch.
This COTS hardware has been selected as it is readily available both to consumers and businesses, and represent an example of a low to mid tier option available at the network edge.

Each of the three less powerful NUCs host an instance of Open5GS \cite{open5gs}, an open-source implementation of a 5G core network written in the C programming language, in one of following configurations.

\begin{itemize}
    \item Bare Metal (BM): in this configuration the 5GC is installed directly on the host OS.
    \item Virtual Machine (VM): in this configuration we make use of the quick emulator (QEMU) hypervisor with kernel-based virtual machine (KVM) acceleration. The 5GC is installed inside a VM running Ubuntu 20.04 Server OS.
    \item Container (CO): in this configuration the 5GC is split into multiple NFs, with each NF running in a separate container on top of the Docker containerization engine.
\end{itemize}

In addition, an instance of Free5GC \cite{free5gc}, another open-source 5GC solution, is deployed on BM as an alternative to Open5GS.
The other two NUCs host the other functions: the 5G RAN, simulated UEs and traffic, monitoring and test case control.
In particular, one NUC is used to run UERANSIM \cite{ueransim}, an open-source state-of-the-art 5G User Equipment (UE) and RAN simulator that can be easily interfaced with Open5GS and Free5GC. For each simulated UE and after the successful establishment of the PDU session, UERANSIM generate a virtual tunneling interface that can be used to send traffic through the network.
For this objective we installed two traffic generator clients, namely Iperf3 and D-ITG \cite{d-itg}.

The fifth and final NUC has a double role. It acts as the end-point for the UE traffic by running the Iperf3 and D-ITG servers. More importantly, it also acts as a controller for the testbed and central collection point for most of the metrics.
The heart of the controller is a Redis database used for its pub/sub functionalities. Multiple Python scripts are deployed in the NUCs to handle various functionalities such as initializing the traffic generator, overseeing the simulated UE connection and collecting selected KPIs.
All these operations require a certain degree of synchronization that is achieved by making the distributed scripts communicate with each other and with the central controller by subscribing and publishing messages on different Redis channels.

\begin{figure}[t]
\centering
\includegraphics[width=\columnwidth]{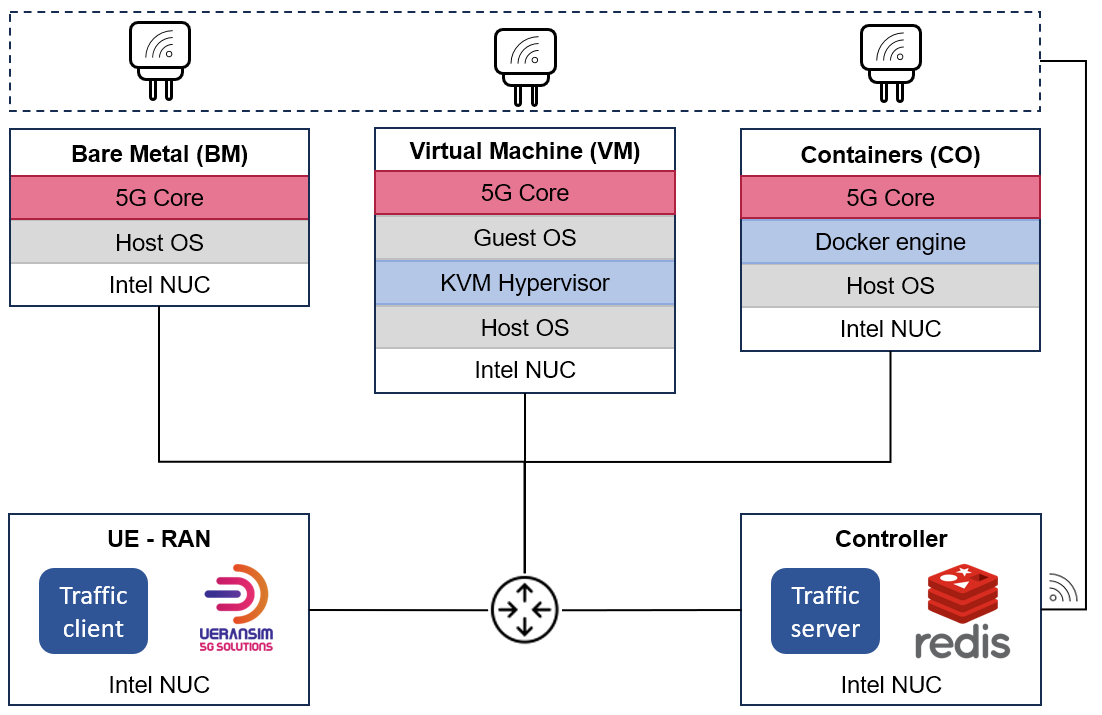}
\caption{Experimental testbed architecture.}
\label{fig:testbed_architecture}
\end{figure}

\subsection{KPI monitoring}
\label{subsec:KPI_monitoring}
With the current implementation of the testbed, we can monitor three types of metrics during the various experiments: power consumption, throughput, and cpu usage.
It is possible to monitor in real-time the power consumption of the 5GC deployments using two different methods, one based on hardware and the other on software as described in Section \ref{subsec:Software-based_power_meters}.
The hardware-based method relies on three Meross MSS310 smart-plugs to which the NUCs are connected. The smart-plugs can be queried via HTTP or MQTT protocols to report the total power consumption of the connected hardware. In our testbed the controller NUC queries the smart-plug data using a specifically crafted HTTP request on a dedicated Wi-Fi network and stores the metrics locally.

As the choice for software-based power meter we opted for Scaphandre v0.5.0 \cite{scaphandre} as it is open-source, easy to deploy and use. The software has two main components: a sensor and an exporter. The sensor collects the power consumption metrics of the host and provide them to exporter who stores them and make them available through various interfaces.
The sensor is based on the Running Average Power Limit (RAPL) interface implemented by most modern Intel CPUs. RAPL is partially based on fully integrated voltage regulators and partially on analytical energy models, but this process is not fully documented and has undergone only limited validation \cite{Desrochers2016}.

To measure the traffic throughput handled by the 5GC deployments, we use the \emph{psutil} Python package \cite{psutil} to directly monitor the NUC network interfaces. The traffic is measured both in terms of bits per seconds and packets per seconds.
Finally, the \emph{psutil} package allows us to also measure the CPU utilization of the NUCs.


\section{Experimental results}
\label{sec:Experimental_results}

In this section we show the results of multiple experiment we carried out with the objective of showcasing the power measurement monitoring methodology in a variety of scenarios and degrees of variability.

\subsection{Artificial CPU load}
\label{subsec:Artificial_CPU_load}

The first experiment is a baseline measurement to validate our setup. The aim is to characterizing how different artificial CPU loads translate into power consumption on the testbed hardware. 
To this end, there are numerous models of varying complexity available in the literature \cite{Makaratzis2017}, however it is not always clear which one is more suited for a specific use-case.
To get first-hand data, we use the \emph{stress-ng} tool  \cite{stress-ng} available on the Ubuntu OS to perform CPU stress tests at various load intervals. At the same time we measure the power consumption registered by the smart-plug connected to the NUC.

\begin{figure}[h]
\centering
\includegraphics[width=\columnwidth]{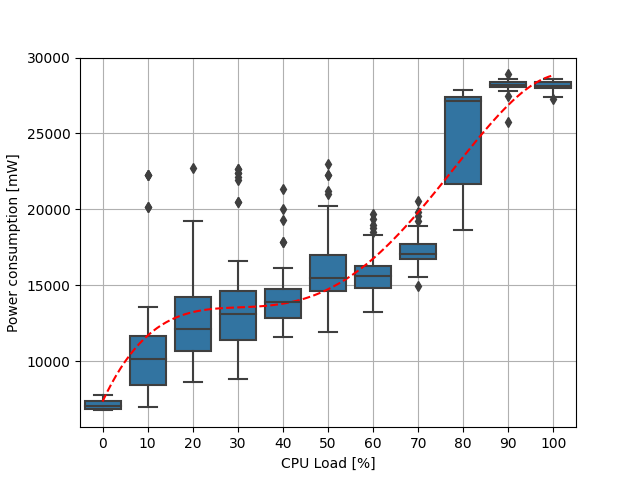}
\caption{Power consumption (in mW) against CPU load (in percentage with respect to the maximum achievable load) measured on the testbed hardware. We interpolated the data with a fourth degree polynomial function showing the nonlinearity}
\label{fig:cpu_power}
\end{figure}

From the results, shown in Figure \ref{fig:cpu_power}, we can notice that the relationship is not linear since it presents the following characteristics.

\begin{itemize}
    \item{0\%-20\%}: rapid increase in power consumption.
    \item{20\%-80\%}: stable or low increase in power consumption on average but high degree of variability.
    \item{80\%-100\%}: high power consumption caused by the Turbo Boost mechanism, which automatically increases the processor's clock operating frequency to keep up with the more demanding tasks at the cost of higher consumption and thermal output.
\end{itemize}

\subsection{Virtualization environments}
\label{subsec:Virtualization_comparison}

\begin{figure*}[!h]
  \begin{center}
  \setlength\belowcaptionskip{-1ex}
    \subfigure[	\label{subfig:smartplug_hist} ]
        {\includegraphics[width=\columnwidth]{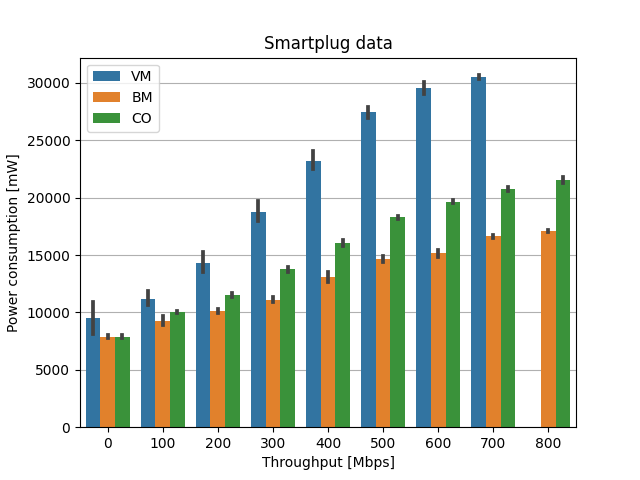}}
    \subfigure[	\label{subfig:scaphandre_hist} ]
        {\includegraphics[width=\columnwidth]{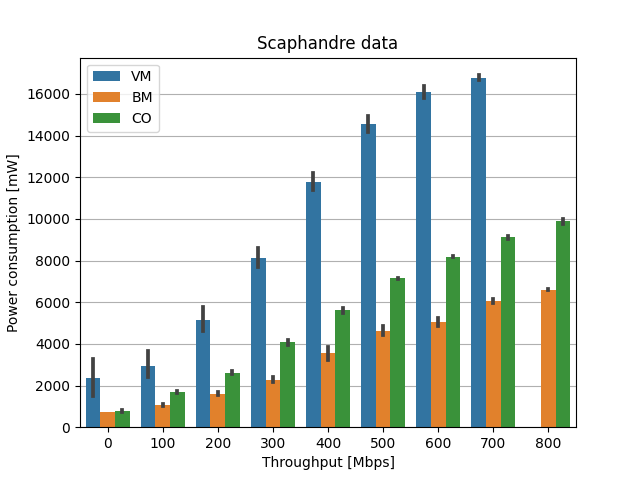}}\\   
    \caption{Power consumption (in mW) for the three different virtualization technologies at different uplink traffic levels. Notice that the VM deployment cannot sustain a throughput higher than 700 Mbps, hence we do not show it past that point.}
    \label{fig:power_hist}
  \end{center}
\end{figure*}

The second and main experiment we conducted is focused on the difference in power consumption of the three virtualization options adopted for the Open5GS deployment. 
We gathered samples every second for a 30 seconds time interval using both the smart-plugs and Scaphandre power meters.
We start the experiment in an idle state, with the 5GC installed and running on the system but with no traffic passing through the network. All consecutive measurements are gathered with increasing levels of up-link traffic generated using \emph{Iperf} and sent across the 5GC user plane.
The simulated traffic varies with increments of $100$~Mbps from $100$~Mbps up to $800$~Mbps, which is the limit of our testbed.
The results are presented in Figure \ref{fig:power_hist}.

Taking first into consideration the smart-plug measurements, shown in Figure \ref{subfig:smartplug_hist}, we can notice that each virtualization option has a characteristic power consumption trend.
Both BM and CO deployments can sustain the maximum tested throughput, while the VM deployment start struggling after the $700$~Mbps mark. For this reason we chose not to show the results for the VM at $800$~Mbps.
As expected, the BM deployment is the most energy efficient since it does not have any virtualization overhead contrary to the other solutions.
The VM deployment, not only cannot sustain a throughput higher than $700$~Mbps, but it also require a significant amount of power, up to $80\%$ more compared to the BM.
On the contrary, the CO deployment proved to be much more energy efficient even with high level of traffic, consuming only $25\%$ more than the BM on the highest traffic scenario.

Similar trends are visible from the Scaphandre measurements shown in Figure \ref{subfig:scaphandre_hist}.
Since the measurements are limited only to the CPU and memory, and not the entire hardware, the difference between the three 5GC deployments are even more accentuated, with the CO and VM deployments consuming up to $50\%$ and $175\%$ more power respectively compared to the BM.

\begin{figure}[t]
\centering
\includegraphics[width=\columnwidth]{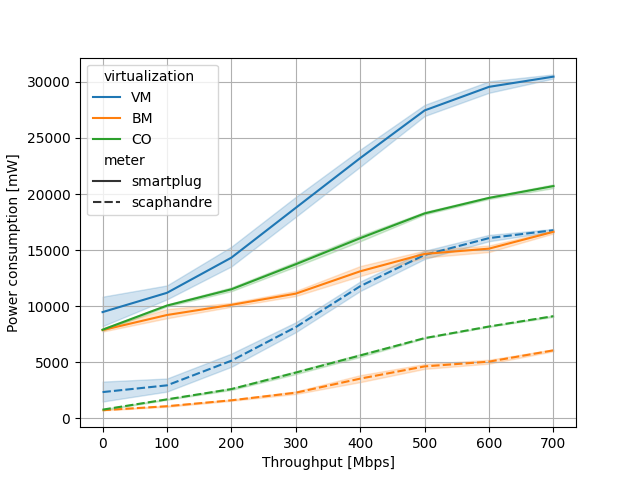}
\caption{Power consumption trends for the 3 different virtualization technologies and 2 power meters.}
\label{fig:scaph_comp}
\end{figure}

From Figure \ref{fig:scaph_comp} we can more clearly see that the BM and CO power consumption trends can be reasonably approximated with a linear function. Meanwhile the VM consumption presents a trend that resemble more a logistic function ranging between the idle power consumption and the maximum hardware power consumption.


\begin{figure}[t]
\centering
\includegraphics[width=\columnwidth]{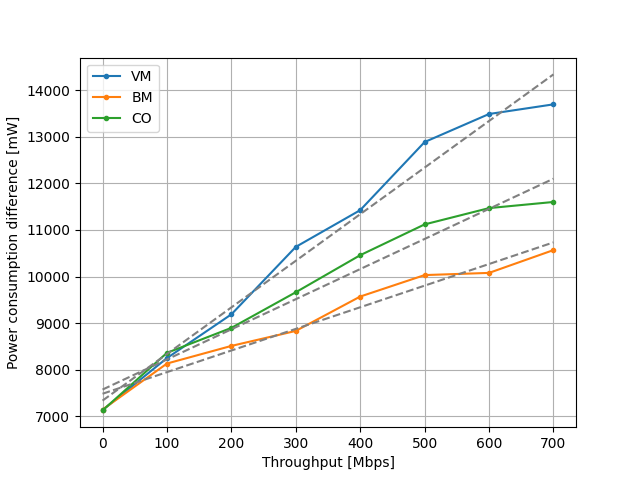}
\caption{Difference between the power measure of the smart-plugs and Scaphandre for the 3 deployments, with linear interpolation.}
\label{fig:scaph_diff}
\end{figure}

The Scaphandre measurements correctly reports the trends but present a noticeable offset compared to the smart plug data. 
This behaviour is consistent with previous studies, such as \cite{Jay2023}. The offsets for the three deployments are graphed in Figure \ref{fig:scaph_diff} together with a linear interpolation. We notice that the offset is not constant but it increases with the throughput with a rate that is different and characteristic to each deployment type.

The additional power measured by the smart-plugs is to be attributed to all the hardware components of the system excluding the CPU and memory, such as hard disk, network interface and power supply. As discussed in Section \ref{subsec:Software-based_power_meters}, this is a common and acknowledged limitation of software-based power meters that operate on top of RAPL interface.
All in all, the relationship between the two alternative power measurements can be described by Equation \ref{eq:1}.

\begin{equation} \label{eq:1}
    P_{hw} = P_{sw} + \alpha \, T + c 
\end{equation}

Where \(P_{hw}\) and \(P_{sw}\) are the power consumption in mW measured by the hardware-based and software-based power meter respectively, \(T\) is the throughput in Mbps,
\(\alpha\) and \(c\) are the coefficients of the linear function specific to the hardware infrastructure and virtualization option.
The values of \(c\) and \(\alpha\) calculated for our experiments on the testbed are given in Table \ref{tab:coef}.

\begin{table}[h]
    \centering
    \begin{tabular}{cccc}
    \toprule
        & VM & BM & CO \\ \midrule
        $\alpha$ & 9.99886 & 4.63977 & 6.46334  \\
        C & 7338  & 7845   & 7575  \\ \bottomrule
    \end{tabular}
    \caption{Coefficients of the linear interpolation.}
    \label{tab:coef}
\end{table}

\subsection{Alternative 5GC implementations}
\label{subsec:Alternative_5GC}
\begin{figure}[t]
\centering
\includegraphics[width=\columnwidth]{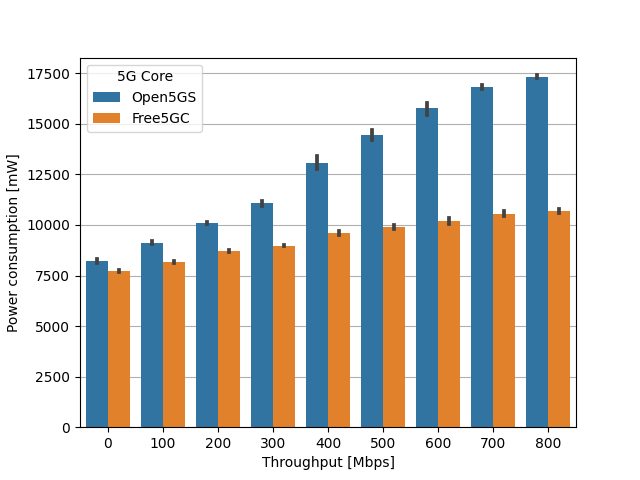}
\caption{Comparison of the power consumption between Open5GS and Free5GC measured using the smart-plugs.}
\label{fig:free5gc_comp}
\end{figure}

In the previous experiments, we have used Open5GS since we consider it simpler to deploy and also more production-ready compared to the other available open-source products.
We now compare the power requirements of an alternative 5GC implementation. For this, we deployed an instance of Free5GC on BM and tested it with the same throughput levels used previously with Open5GS. The smart-plug power metrics are shown in Figure \ref{fig:free5gc_comp}.

We can notice that Free5GC consumes substantially less power than Open5GS, almost $40\%$ less at 800 Mbps. Such a big difference is mainly due to the different packet processing techniques adopted. While Open5GS handles the packet processing in the User Space, Free5GC uses Kernel Space and requires the GTP5G customized Linux kernel module \cite{Free5GC:gtp5g}.
This characteristic provides better performance but it adds an additional level of complexity and can limit the type of infrastructure on which the core can be deployed.
Additional information and a more thorough comparison between User Space and In-Kernel packet processing can be found in \cite{Parola2023}.

\subsection{Per-process consumption}
\label{subsec:Processes_consumption}

\begin{figure*}[h]
  \begin{center}
  \setlength\belowcaptionskip{-1ex}
    \subfigure[	\label{subfig:proc_ogs} ]
        {\includegraphics[width=\columnwidth]{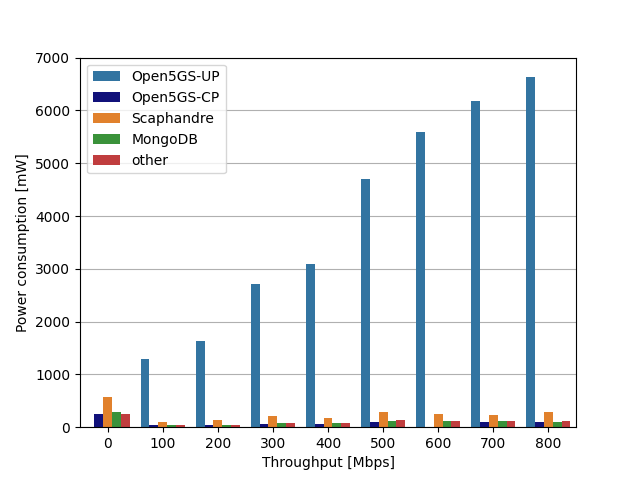}}
    \subfigure[	\label{subfig:proc_fgc} ]
        {\includegraphics[width=\columnwidth]{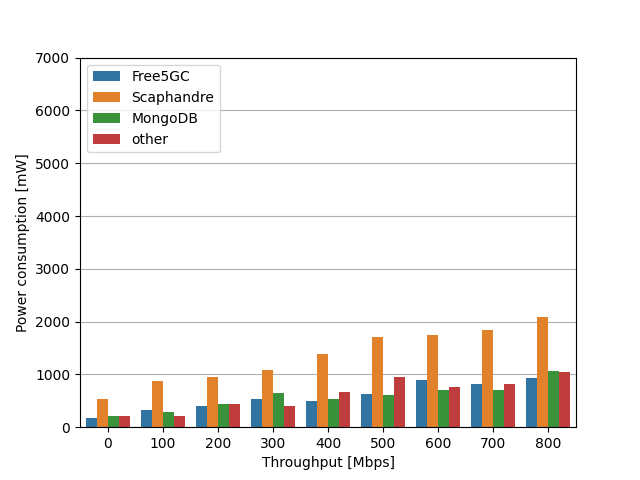}}\\   
    \caption{Power consumption measurements of selected processes measured using Scaphandre with the Open5GS core in (a) and Free5GC in (b). Note that the measurements present some inaccuracies.}
    \label{fig:proc_bars}
  \end{center}
\end{figure*}

In most practical use cases, it is useful to understand not only the power consumption of the overall 5GC deployment but also of the single NFs. Such knowledge can be leveraged to implement better scaling and orchestration mechanisms aimed directly at single NFs.
In particular, when talking about 5G and edge computing, it is often only the User Plane Function (UPF) that is deployed on-premise to improve latency and privacy.
Scaphandre allows for the monitoring of the power consumption of single processes by matching the instances in which each process is being allocated CPU time to the corresponding power readings. By cleverly aggregating this data, Scaphandre provides a reasonable estimation even when dealing with multiple processors and virtualized environments.

We present the results we obtained for both Open5GS and Free5GC in Figure \ref{subfig:proc_ogs} and \ref{subfig:proc_fgc} respectively.
We highlight the consumption of the selected 5GC, the MongoDB database and of Scaphandre itself.
The aggregate power consumption of all other processes is listed as \emph{other}.
It is important to note that in both the Open5GS and Free5GC software, each NF runs as a different process. In the case of Free5GC we grouped the power consumption of all components of the core under the voice \emph{Free5GC} as they would be too low individually.
Conversely, in the Open5GS data we singled out the consumption of the UPF and marked it as \emph{Open5GS-UP} while grouping all other NFs under the tag \emph{Open5GS-CP}. 
The \emph{MongoDB} database is mainly used by the control-plane functions to store subscribers data and other management informations, so it can be considered part of the 5GC even if we decided to keep it separate in the plots.

Before analyzing the results, we have to point out that we noticed some inaccuracies and inconsistencies in the measurements. 
In particular, in the Open5GS deployment (Figure \ref{subfig:proc_ogs}) the power consumption of all other processes get significantly lower when the UPF consumption starts increasing. In our opinion this is due to a wrong estimation made by Scaphandre when a single process has a much higher consumption with respect to all others. On the contrary, in the Free5CG deployment (Figure \ref{subfig:proc_fgc}) the consumption of all processes increases with the throughput. It is reasonable to assume that this is also due to an imprecise estimation when handling workloads that are not CPU-intensive.
There have been reports of errors and documented inaccuracies in Scaphandre process monitoring mechanism, especially when measuring idle hardware and RAM intensive workloads but substantial improvements are planned for the release of Scaphandre version 1.0.
Nevertheless, we think that this mechanism could prove very useful if further developed and that the gathered metrics can provide some helpful insights.

In particular, we can infer the two following observations:
\begin{itemize}
    \item In the Open5GS deployment, the UPF has a much higher power consumption with respect to all other processes. On the contrary, the consumption of the UPF in Free5GC is negligible since the packet processing is handle by the GTP5G kernel module, as explained in Section \ref{subsec:Alternative_5GC}.
    \item Scaphandre itself registers a small but significant power consumption among all scenarios. This is an acknowledged drawback of, in a higher or lower measure, all software-based power meters.
    In our specific study, the main culprit is probably the process monitoring component since it requires much heavier real-time calculations compared to the RAPL sensor. 
\end{itemize}

\subsection{Multiple UEs}
\label{subsec:Multiple_UEs}

\begin{figure}[h]
\centering
\includegraphics[width=\columnwidth]{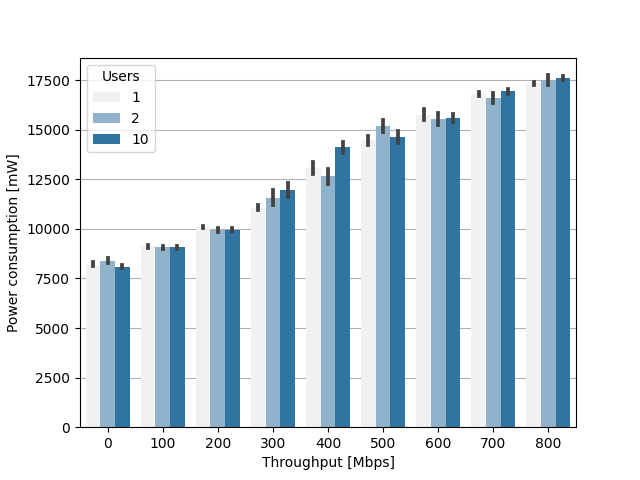}
\caption{Power consumption of the BM deployment with multiple UEs connected measured using the smart-plugs.}
\label{fig:multi_ue_bars}
\end{figure}

\begin{figure}[h]
\centering
\includegraphics[width=\columnwidth]{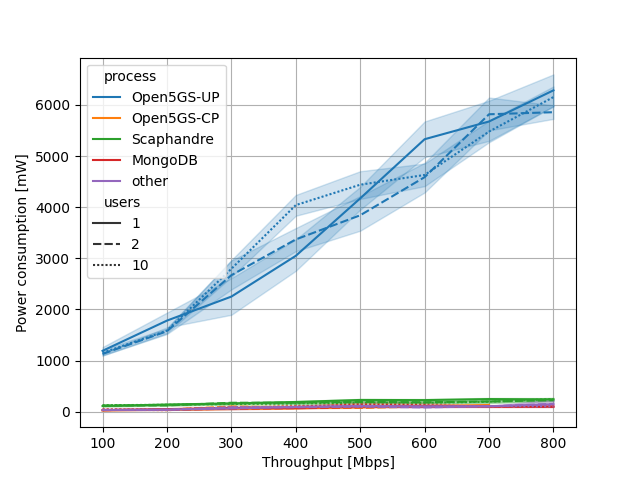}
\caption{Power consumption of the processes in the BM deployment with multiple UEs connected measured using Scaphandre.}
\label{fig:multi_ue_proc}
\end{figure}

Up until now, all experiments have been carried out by simulating a single UE through which all the traffic is sent. 
In Figure \ref{fig:multi_ue_bars} we show the results of power consumption measurement using the smart-plugs on the BM deployment when multiple UEs are connected. The overall throughput is equally divided among all UEs so that we can fairly compare scenarios such as 1 UE transmitting at $100$~Mbps and 10 UEs transmitting at $10$~Mbps each.
We compare the measured power with 1, 2 and 10 UEs connected. The power consumption measured by the smart-plug does not show any substantial variation outside of the confidence interval.
The same can be observed from the per-process measurement gather with Scaphandre and shown in Figure \ref{fig:multi_ue_proc}.

Due to the nature of our testbed and the limited hardware specifications, it is quite difficult to simulate a scenario with a higher number of connected UEs. 
Furthermore, few software solutions capable of simulating realistic subscriber behavior and complex control plane traffic are available in the market. 
It is possible that in particular scenarios, such as in the case of massive IoT where a high number of devices are connected to the network but requiring only low throughput, the results could be significantly different from the one reported in this study.

\section{Lessons learned}

This section aims at analyzing the achieved results and put them in perspective
relating the evolution of 5G SBA towards 6G and beyond.

First of all, the reader should remember that the scope of this paper is not
to derive a generalist mathematical power consumption model, but rather to underline
(as results properly demonstrate) that power consumption of the 5GC can be highly
variable, based on the selected hardware, the adopted virtualization architecture,
and the traffic and processing load. Indeed, the paper aims at providing an
approach to make power consumption of 5GC observable that is agnostic to the hardware infrastructure, virtualization and software implementation.

Such an approach can be leveraged in various ways by mobile network operators and other involved stakeholders to implement green orchestration policies, train machine learning algorithms, reduce the operating expenses or gain additional awareness of the 5GC deployment's energy requirements. The ability to expose the energy footprint and real-time metrics not only to internal management tools, but also to all stakeholders involved represents a key step in achieving the sustainability targets envisioned in future 6G networks \cite{6Green}. 

From a quantitative viewpoint, as expected, bare metal deployment provides a significantly more efficient solution than the other virtualized environments.
However, other metrics and aspects must be considered in the implementation and management of a 5GC deployment. Such points are briefly summarized in the following paragraphs.

In general, containerization provides a superior degree of agility and scalability, characterized by its lightweight nature, portability, and seamless deployment capabilities. It enables the establishment of a continuous integration and continuous deployment pipeline, thereby reducing the time required to introduce new network functionalities and features to the market. However, the management of containers and their dependencies may raise intricate challenges. Furthermore, containerization represents a relatively recent technology requiring substantial effort for integration with legacy or pre-existing solutions, leading many operators to continue relying on time-tested and reliable VM-based infrastructures. VMs also provide a higher level of isolation compared to containers, even if not comparable with the physical isolation of bare metal. The drawback is that a bare metal deployment requires the full ownership and control of the physical server infrastructure that must be dedicated only to the 5GC.
Moreover, the selection of one solution is based on constraints or decisions related to the ownership level of the considered scenario. 
Indeed, from an operator viewpoint, the ownership level (and related management costs) of the mobile network infrastructure radically reduces when transitioning from legacy bare metal to more sustainable and scalable solutions such as VMs and containers, as it allows multiple mobile operators to share network infrastructures and management costs to enhance service coverage and reduce the overall costs of the network deployment.   
The above considerations can be summarized in Table \ref{tab:virt}, where a qualitative comparison summarizing the main features of the three considered virtualization environments is presented. Each virtualization solution is classified, based on achieved results and technical approach, in terms of achievable energy efficiency, scalability level, security, and ownership level.

\begin{table}[h]
    \centering
    \begin{tabular}{c|ccc}
    \toprule
        & VM & BM & CO \\ \midrule
        energy efficiency & low & high & mid  \\
        scalability & mid  & low   & high  \\
        security & mid & high & low  \\
        ownership level & mid & high & low \\\bottomrule
    \end{tabular}
    \caption{A qualitative comparison of most relevant virtualization environments.}
    \label{tab:virt}
\end{table}

\section{Conclusions}
\label{sec:Conclusions}

In a context where the importance of energy efficiency for networks is growing, this paper analyzes the issue of real-time power consumption monitoring in heterogeneous 5GC deployments.
We propose a measurement-based approach and validate it on a testbed built with COTS hardware and open-source software simulating an edge computing environment. We considered two 5GC solutions (Open5GS and Free5GC) and three alternative virtualization options, namely bare metal, virtual machines and containers.
Results show that a deployment based on virtual machines might require up to $80\%$ more energy compared to the bare metal, while the containerized deployment only $25\%$ more.
When analyzing the 5GC in a data-plane traffic intensive scenario, we found that, as expected, the UPF is the main component responsible for the power consumption. In this scenario, Free5GC presents a much lower power consumption compared to Open5GS thanks to its in-kernel packet processing mechanism.
The power consumption metrics are simultaneously gathered from hardware-based and software-based meters, allowing us to model the differences between the two by using a linear function with deployment-specific coefficients.

In the future, we plan to integrate this monitoring system in a wider network orchestration platform. This will allow the gathered metrics to be leveraged in the design and implementation of green policies and optimization algorithms for the OAM of both public and private 5GC networks. 
In particular, we will use the gathered data to train a reinforcement learning algorithm capable of optimizing the energy efficiency of an 5GC deployment in an edge to cloud continuum environment.

\section{Data availability}
\label{sec:data}
The data gathered from the experimental testbed and used in this work is available on GitHub \cite{github_data}.

\section{Acknowledgments}
\label{sec:Acknowledgments}
This work is an extension of the conference paper \cite{Bellin2023}.

The doctoral studies and research work of Arturo Bellin are jointly supported by Athonet, a HPE acquisition, and the Italian National Inter-University Consortium for Telecommunications (CNIT).
The work leading to this publication was partially funded by the European Union’s Horizon Europe under Grant Agreement no. 101096342 (HORSE project) and Grant Agreement no. 101096925 (6Green project).


\bibliographystyle{elsarticle-num} 
\bibliography{bibliography}

\begin{thebibliography}{10}
\expandafter\ifx\csname url\endcsname\relax
  \def\url#1{\texttt{#1}}\fi
\expandafter\ifx\csname urlprefix\endcsname\relax\def\urlprefix{URL }\fi
\expandafter\ifx\csname href\endcsname\relax
  \def\href#1#2{#2} \def\path#1{#1}\fi

\bibitem{ericsson:ran}
\href{https://www.ericsson.com/en/blog/2021/12/a-holistic-approach-to-address-ran-energy-efficiency}{A
  holistic approach to address ran energy efficiency}, Tech. rep., Ericsson
  (2021).
\newline\urlprefix\url{https://www.ericsson.com/en/blog/2021/12/a-holistic-approach-to-address-ran-energy-efficiency}

\bibitem{NRDC:datacenter}
\href{https://www.nrdc.org/sites/default/files/data-center-efficiency-assessment-IP.pdf}{Data
  center efficiency assessment—scaling up energy efficiency across the data
  center industry:evaluating key drivers and barriers}, Tech. rep., Natural
  Resources Defense Council (2014).
\newline\urlprefix\url{https://www.nrdc.org/sites/default/files/data-center-efficiency-assessment-IP.pdf}

\bibitem{ITU:carbon}
\href{https://www.itu.int/rec/T-REC-L.1333-202209-I}{L.1333: Carbon data
  intensity for network energy performance monitoring}, Tech. rep.,
  International Telecommunication Union Telecommunication Standardization
  Sector (2022).
\newline\urlprefix\url{https://www.itu.int/rec/T-REC-L.1333-202209-I}

\bibitem{ETSI:ZSM}
ETSI GS ZSM 002,
  \href{https://www.etsi.org/deliver/etsi_gs/ZSM/001_099/002/01.01.01_60/gs_ZSM002v010101p.pdf}{Zero-touch
  network and Service Management (ZSM); Reference Architecture}, v1.1.1 (Aug
  2019).
\newline\urlprefix\url{https://www.etsi.org/deliver/etsi_gs/ZSM/001_099/002/01.01.01_60/gs_ZSM002v010101p.pdf}

\bibitem{Depasquale2023}
E.-V. Depasquale, F.~Davoli, H.~Rajput,
  \href{https://www.mdpi.com/1424-8220/23/1/255}{Dynamics of research into
  modeling the power consumption of virtual entities used in the telco cloud},
  Sensors 23~(1) (2023).
\newblock \href {https://doi.org/10.3390/s23010255}
  {\path{doi:10.3390/s23010255}}.
\newline\urlprefix\url{https://www.mdpi.com/1424-8220/23/1/255}

\bibitem{Jiang2019}
C.~Jiang, Y.~Wang, D.~Ou, Y.~Li, J.~Zhang, J.~Wan, B.~Luo, W.~Shi,
  \href{https://www.sciencedirect.com/science/article/pii/S2210537917300963}{Energy
  efficiency comparison of hypervisors}, Sustainable Computing: Informatics and
  Systems 22 (2019) 311--321.
\newblock \href {https://doi.org/https://doi.org/10.1016/j.suscom.2017.09.005}
  {\path{doi:https://doi.org/10.1016/j.suscom.2017.09.005}}.
\newline\urlprefix\url{https://www.sciencedirect.com/science/article/pii/S2210537917300963}

\bibitem{Morabito2015}
R.~Morabito, Power consumption of virtualization technologies: An empirical
  investigation, in: 2015 IEEE/ACM 8th International Conference on Utility and
  Cloud Computing (UCC), 2015, pp. 522--527.
\newblock \href {https://doi.org/10.1109/UCC.2015.93}
  {\path{doi:10.1109/UCC.2015.93}}.

\bibitem{Shea2014}
R.~Shea, H.~Wang, J.~Liu, Power consumption of virtual machines with network
  transactions: Measurement and improvements, in: IEEE INFOCOM 2014 - IEEE
  Conference on Computer Communications, 2014, pp. 1051--1059.
\newblock \href {https://doi.org/10.1109/INFOCOM.2014.6848035}
  {\path{doi:10.1109/INFOCOM.2014.6848035}}.

\bibitem{Behravesh2019}
R.~Behravesh, E.~Coronado, R.~Riggio, Performance evaluation on virtualization
  technologies for nfv deployment in 5g networks, in: 2019 IEEE Conference on
  Network Softwarization (NetSoft), 2019, pp. 24--29.
\newblock \href {https://doi.org/10.1109/NETSOFT.2019.8806664}
  {\path{doi:10.1109/NETSOFT.2019.8806664}}.

\bibitem{Aggarwal2020}
V.~Aggarwal, B.~Thangaraju, Performance analysis of virtualisation technologies
  in nfv and edge deployments, in: 2020 IEEE International Conference on
  Electronics, Computing and Communication Technologies (CONECCT), 2020, pp.
  1--5.
\newblock \href {https://doi.org/10.1109/CONECCT50063.2020.9198367}
  {\path{doi:10.1109/CONECCT50063.2020.9198367}}.

\bibitem{Adoga2022}
H.~U. Adoga, Y.~Elkhatib, D.~P. Pezaros, On the performance benefits of
  heterogeneous virtual network function execution frameworks, in: 2022 IEEE
  8th International Conference on Network Softwarization (NetSoft), 2022, pp.
  109--114.
\newblock \href {https://doi.org/10.1109/NetSoft54395.2022.9844115}
  {\path{doi:10.1109/NetSoft54395.2022.9844115}}.

\bibitem{Reddy2023}
R.~Reddy, M.~Gundall, C.~Lipps, H.~Schotten, Open source 5g core network
  implementations: A qualitative and quantitative analysis, 2023.

\bibitem{Lando2023}
G.~Lando, L.~A.~F. Schierholt, M.~P. Milesi, J.~A. Wickboldt, Evaluating the
  performance of open source software implementations of the 5g network core,
  in: NOMS 2023-2023 IEEE/IFIP Network Operations and Management Symposium,
  2023, pp. 1--7.
\newblock \href {https://doi.org/10.1109/NOMS56928.2023.10154399}
  {\path{doi:10.1109/NOMS56928.2023.10154399}}.

\bibitem{Bellin2023}
A.~Bellin, M.~Centenaro, F.~Granelli, A preliminary study on the power
  consumption of virtualized edge 5g core networks, in: 2023 IEEE 9th
  International Conference on Network Softwarization (NetSoft), 2023, pp.
  420--425.
\newblock \href {https://doi.org/10.1109/NetSoft57336.2023.10175489}
  {\path{doi:10.1109/NetSoft57336.2023.10175489}}.

\bibitem{ts:28.310}
The 3GPP Association,
  \href{https://www.etsi.org/deliver/etsi_gs/ZSM/001_099/002/01.01.01_60/gs_ZSM002v010101p.pdf}{TS
  28.310 Management and orchestration; Energy efficiency of 5G}, v17.5.0 (May
  2023).
\newline\urlprefix\url{https://www.etsi.org/deliver/etsi_gs/ZSM/001_099/002/01.01.01_60/gs_ZSM002v010101p.pdf}

\bibitem{6Green}
\href{https://www.6green.eu/}{Green technologies for 5/6g service-based
  architectures. deliverable d2.1 use and business cases, design and technology
  requirements, and architecture specification}.
\newline\urlprefix\url{https://www.6green.eu/}

\bibitem{Jay2023}
M.~Jay, V.~Ostapenco, L.~Lefevre, D.~Trystram, A.-C. Orgerie, B.~Fichel, An
  experimental comparison of software-based power meters: focus on cpu and gpu,
  in: 2023 IEEE/ACM 23rd International Symposium on Cluster, Cloud and Internet
  Computing (CCGrid), 2023, pp. 106--118.
\newblock \href {https://doi.org/10.1109/CCGrid57682.2023.00020}
  {\path{doi:10.1109/CCGrid57682.2023.00020}}.

\bibitem{Ismail2020}
L.~Ismail, H.~Materwala, \href{https://doi.org/10.1145/3390605}{Computing
  server power modeling in a data center: Survey, taxonomy, and performance
  evaluation}, ACM Comput. Surv. 53~(3) (jun 2020).
\newblock \href {https://doi.org/10.1145/3390605} {\path{doi:10.1145/3390605}}.
\newline\urlprefix\url{https://doi.org/10.1145/3390605}

\bibitem{open5gs}
\href{https://open5gs.org/}{Open5gs}, accessed: October~2023.
\newline\urlprefix\url{https://open5gs.org/}

\bibitem{free5gc}
\href{https://free5gc.org/}{Free5gc}.
\newline\urlprefix\url{https://free5gc.org/}

\bibitem{ueransim}
\href{https://github.com/aligungr/UERANSIM}{Ueransim}.
\newline\urlprefix\url{https://github.com/aligungr/UERANSIM}

\bibitem{d-itg}
\href{https://traffic.comics.unina.it/software/ITG/}{D-itg}.
\newline\urlprefix\url{https://traffic.comics.unina.it/software/ITG/}

\bibitem{scaphandre}
\href{https://github.com/hubblo-org/scaphandre}{Scaphandre}.
\newline\urlprefix\url{https://github.com/hubblo-org/scaphandre}

\bibitem{Desrochers2016}
S.~Desrochers, C.~Paradis, V.~M. Weaver,
  \href{https://doi.org/10.1145/2989081.2989088}{A validation of dram rapl
  power measurements}, in: Proceedings of the Second International Symposium on
  Memory Systems, MEMSYS '16, Association for Computing Machinery, New York,
  NY, USA, 2016, p. 455–470.
\newblock \href {https://doi.org/10.1145/2989081.2989088}
  {\path{doi:10.1145/2989081.2989088}}.
\newline\urlprefix\url{https://doi.org/10.1145/2989081.2989088}

\bibitem{psutil}
\href{https://psutil.readthedocs.io/en/latest/}{Psutil}.
\newline\urlprefix\url{https://psutil.readthedocs.io/en/latest/}

\bibitem{Makaratzis2017}
A.~T. Makaratzis, C.~K. Filelis-Papadopoulos, K.~M. Giannoutakis, G.~A.
  Gravvanis, D.~Tzovaras, \href{https://doi.org/10.1145/3139367.3139409}{A
  comparative study of cpu power consumption models for cloud simulation
  frameworks}, in: Proceedings of the 21st Pan-Hellenic Conference on
  Informatics, PCI '17, Association for Computing Machinery, New York, NY, USA,
  2017.
\newblock \href {https://doi.org/10.1145/3139367.3139409}
  {\path{doi:10.1145/3139367.3139409}}.
\newline\urlprefix\url{https://doi.org/10.1145/3139367.3139409}

\bibitem{stress-ng}
\href{https://wiki.ubuntu.com/Kernel/Reference/stress-ng}{Stress-ng}.
\newline\urlprefix\url{https://wiki.ubuntu.com/Kernel/Reference/stress-ng}

\bibitem{Free5GC:gtp5g}
\href{https://free5gc.org/blog/20230920/Introduction_of_gtp5g_and_some_kernel_concepts/}{Introduction
  of gtp5g and some kernel concepts}.
\newline\urlprefix\url{https://free5gc.org/blog/20230920/Introduction_of_gtp5g_and_some_kernel_concepts/}

\bibitem{Parola2023}
F.~Parola, R.~Procopio, R.~Querio, F.~Risso,
  \href{https://doi.org/10.1145/3594255.3594257}{Comparing user space and
  in-kernel packet processing for edge data centers}, SIGCOMM Comput. Commun.
  Rev. 53~(1) (2023) 14–29.
\newblock \href {https://doi.org/10.1145/3594255.3594257}
  {\path{doi:10.1145/3594255.3594257}}.
\newline\urlprefix\url{https://doi.org/10.1145/3594255.3594257}

\bibitem{github_data}
\href{https://github.com/IncludeArthur/5GC-power-consumption-data}{5gc power
  consumption data}.
\newline\urlprefix\url{https://github.com/IncludeArthur/5GC-power-consumption-data}

\end{thebibliography}





\end{document}